\documentclass[twocolumn]{emulateapj}

\begin{document}

\title{Optical and Dynamical Characterization of Comet-Like Main-Belt Asteroid (596) Scheila
        \footnote{
        Some of the data presented herein were obtained at the W.\ M.\
        Keck Observatory, which is operated as a scientific partnership
        among the California Institute of Technology, the University of
        California, and the National Aeronautics and Space Administration.
        The Observatory was made possible by the generous financial support
        of the W.\ M.\ Keck Foundation.
        }
}

\author{Henry H. Hsieh$^{a,b}$, Bin Yang$^{a}$, Nader Haghighipour$^{a}$}

\affil{
    $^{a}$Institute for Astronomy, Univ. of Hawaii, 2680 Woodlawn Drive, Honolulu, Hawaii 96822, USA\newline
    $^{b}$Hubble Fellow
}
\email{hsieh@ifa.hawaii.edu, yangbin@ifa.hawaii.edu, nader@ifa.hawaii.edu}

\slugcomment{Submitted, 2011-05-09; Accepted, 2011-09-14}

\begin{abstract}
We present observations and a dynamical analysis of the comet-like main-belt object,
(596) Scheila.  $V$-band photometry obtained on UT 2010 December 12 indicates that
Scheila's dust cloud has a scattering
cross-section $\sim1.4$ times larger than that of the nucleus, corresponding
to a dust mass of $M_d\sim3\times10^7$~kg.  $V-R$ color measurements indicate that
both the nucleus and dust are redder than the Sun, with no significant color differences between
the dust cloud's northern and southern plumes.  We also undertake an ultimately unsuccessful
search for CN emission, where we find CN and H$_2$O production rates of
$Q_{\rm CN}<9\times10^{23}$~s$^{-1}$ and $Q_{\rm H_2O}<10^{27}$~s$^{-1}$.  Numerical
simulations indicate that Scheila is dynamically stable for $>100$~Myr, suggesting that it is likely
native to its current location. We also find that it does not belong to a dynamical asteroid family of any
significance.  We consider sublimation-driven scenarios that could produce
the appearance of multiple plumes of dust emission, but reject them as being physically implausible.
Instead, we concur with previous studies that the unusual morphology of Scheila's dust cloud is
most simply explained by a single oblique impact, meaning this object is likely not a main-belt comet, but
is instead the second disrupted asteroid after P/2010 A2 (LINEAR) to be discovered.
\end{abstract}

\keywords{comets: general ---
          minor planets, asteroids}

\newpage

\section{INTRODUCTION}
\label{intro}

Discovered on 1906 February 21, main-belt asteroid (596) Scheila has
a diameter of $d=113.34$~km \citep{ted04}, an orbital period of $P_{\rm orb}=5.01$~years,
a semimajor axis of $a=2.927$~AU,
an eccentricity of $e=0.165$, an inclination of
$i=14.66^{\circ}$, and a Tisserand parameter (with respect
to Jupiter) of $T_J=3.209$.  On 2010 December 10.4, observations by S.\ Larson using the 0.68~m
Catalina Schmidt telescope showed that Scheila was exhibiting comet-like activity.
Examination of archived Catalina data indicated
that the object also appeared diffuse on December 3, but was
point-source-like
on October 18, November 2, and November 11 \citep{lar10}.

\begin{figure}
\plotone{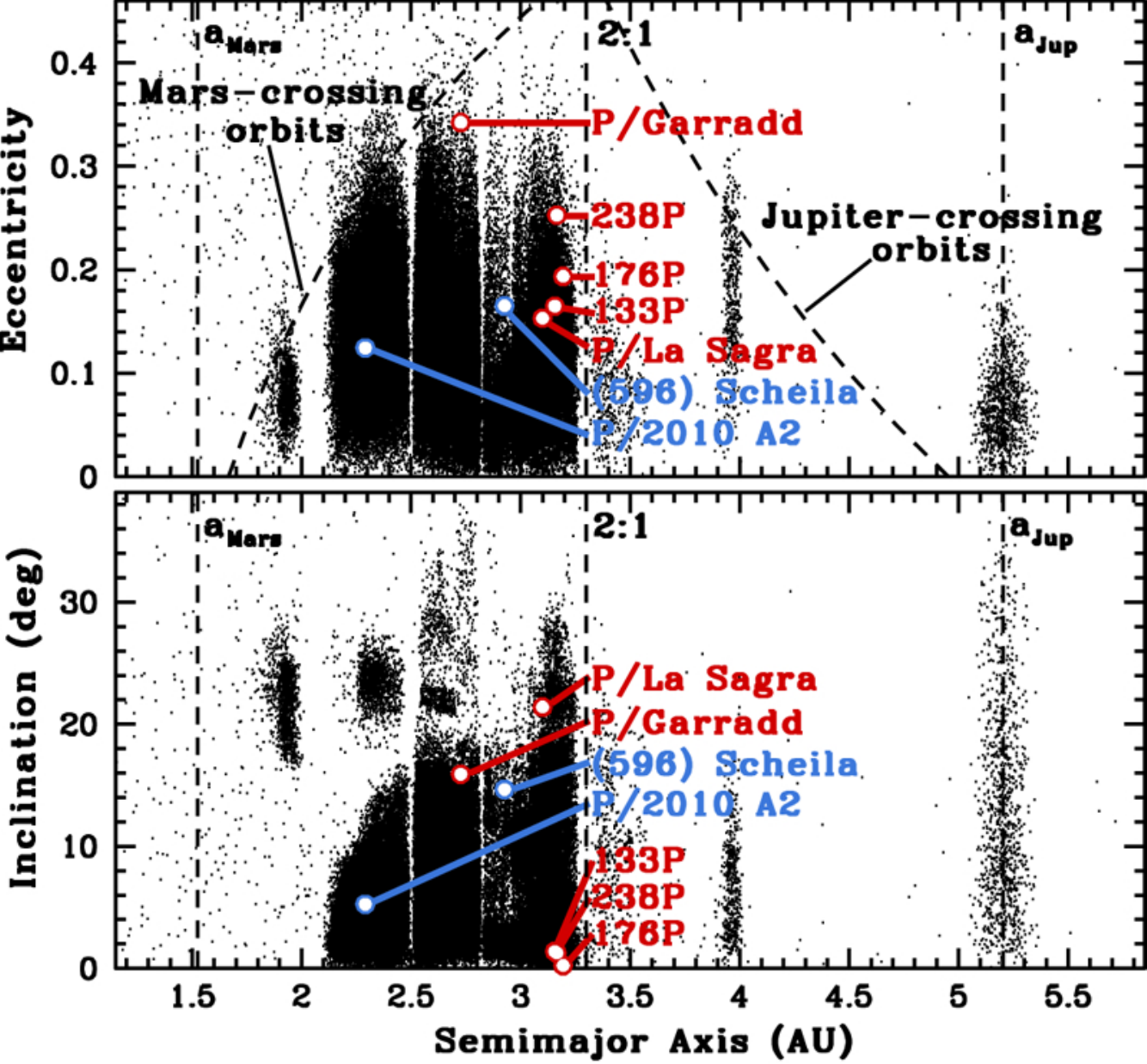}
\caption{\small 
Plots of eccentricity (top) and inclination (bottom)
versus semimajor axis showing the distributions in orbital element space of main-belt asteroids (black dots),
main-belt comets (red circles), and likely disrupted asteroids P/2010 A2 and Scheila (blue circles).
Also marked with dotted lines are the semimajor axes of Mars ($a_{\rm Mars}$)
and Jupiter ($a_{\rm Jup}$),
the semimajor axis of the 2:1 mean-motion resonance with Jupiter, and the loci of Mars-crossing orbits and Jupiter-crossing
orbits.
}
\label{mbcs_aei}
\end{figure}

If its activity is truly cometary, Scheila would be the newest member of the small
class of objects known as main-belt comets (MBCs),
which are
objects that exhibit cometary activity ({\it e.g.}, dust emission) due to the
sublimation of volatile ices, but occupy stable orbits entirely confined to the
main asteroid belt (Fig.~\ref{mbcs_aei}).
MBCs are unlikely to originate in the outer solar system like other comets,
and are probably native to the main belt \citep{jfer02,hag09}.  Their activity is believed to be triggered by
impacts that excavate subsurface volatile material \citep[probably water ice;][]{pri09,sch08}, exposing it to
direct solar heating, driving sublimation and comet-like dust emission \citep{hsi06}.
Noting the strong correlation between activity and proximity to perihelion for all currently known MBCs,
however, \citet{hsi11b} proposed the alternate possibility that activity is in fact primarily dependent
on heliocentric distance, and only peaks during the post-perihelion portion of each MBC's orbit due to
the finite time required for solar thermal waves to propagate through insulating surface material before
reach the subsurface reservoirs of volatile material below.  Currently, both hypotheses are consistent with
the available evidence, though we note the first hypothesis could be ruled out (at least as a universally
applicable explanation for MBC activity modulation) by the determination of a pole orientation for an MBC that
is inconsistent with seasonal modulation of activity ({\it i.e.}, where peak activity does not occur near a
solstice position, as required by the seasonal modulation hypothesis, but near an equinox position).  Likewise,
the second hypothesis could be ruled out as a universal explanation for MBC activity modulation by the discovery
of an MBC which exhibits peak activity prior to reaching perihelion.

To date, gas emission has never been directly detected
for an MBC \citep[{\it e.g.},][]{jew09,lic11b}. This present lack of spectroscopic confirmation of
outgassing is largely due to the faintness of MBC activity,
requiring extremely sensitive observations.
Gas detection efforts are further complicated by the transience of MBC activity, requiring that
spectroscopic observations be secured
soon after new MBCs are discovered, something that has only been achieved for the more
recent discoveries.  Since rapid gas dissipation
timescales mean that dust emission may continue long after active sublimation has actually ceased,
even those observations may not have been sufficiently prompt to detect gas.

Despite the lack of direct gas detections, volatile sublimation is
inferred to be the driver of activity for all of the objects considered to be MBCs.
The indirect evidence for cometary
activity typically includes observations showing
recurrent activity and dust modeling showing continuous dust emission during each active episode, both consistent only
with sublimation-driven dust ejection \citep{hsi04,hsi11a,hsi11b}.

There is, however, the case of  P/2010 A2 (LINEAR), which orbits in the main belt and exhibited a comet-like
dust tail in 2010.  \citet{mor10} initially reported that numerical dust modeling indicated that the object's
apparent activity was due to emission that persisted over several months, implying that it was likely due to
ice sublimation.  However, follow-up modeling based on high-resolution Wide Field Camera 3 images from the
Hubble Space Telescope \citep{jew10} found that the apparent activity of P/2010 A2 was actually most likely a
consequence of an inter-asteroid impact, rather than the product of sublimation.
Observations revealed a gap between P/2010 A2's dust tail and its nucleus, suggesting that the dust was ejected
in a single burst, as would be expected from an impact-generated ejecta cloud, and was drifting away.
This scenario was corroborated with further modeling by \citet{sno10}, who employed both ground-based data
and data from the OSIRIS Narrow Angle Camera aboard the European Space Agency's Rosetta spacecraft, and thus
were able to study the dust tail from two different viewing perspectives, providing additional constraints on
their model.  Thus, while P/2010 A2 may have
appeared comet-like, it is likely not a true comet ({\it i.e.}, an object that exhibits sublimation-driven
activity) and is better characterized as a disrupted asteroid, that is, an object that exhibits dust
emission as a consequence of an impact (or impacts), and not through the action of sublimating ice.

All studies of Scheila's 2010 outburst published to date \citep{jew11,bod11,yan11,ish11a,ish11b,mor11b}
conclude that, like for P/2010 A2, Scheila's dust cloud is likely be impact-generated, and not a consequence of
sublimation.
In this manuscript,
we describe the results of work aimed at better understanding the full nature of Scheila and its
2010 outburst that
provide additional evidence supporting this conclusion.

\section{OBSERVATIONS\label{observations}}

We observed Scheila on multiple occasions
shortly after the discovery of its comet-like activity.
Optical imaging was performed in photometric conditions on UT 2010 December 12 using the University of Hawaii (UH) 2.2~m
telescope, while optical spectroscopy was obtained using the 10~m Keck I telescope on UT 2010 December 17.
Details of these observations are listed in Table~\ref{obslog}.
Results of additional near-infrared observations conducted as part of this observational campaign are
reported in \citet{yan11}.

Imaging observations employed a Tektronix 2048$\times$2048 pixel CCD ($0\farcs219$ pixel$^{-1}$)
behind Kron-Cousins filters.
Spectroscopic observations were made using Keck's Low Resolution Imaging
Spectrometer \citep[LRIS;][]{oke95}, a two-channel instrument that splits incoming light into red and
blue components, and employs dual Tektronix 2048$\times$2048 CCDs ($0\farcs135$ pixel$^{-1}$).
We used a $1\farcs0$-wide long-slit mask, 600/4000 grism, and 560 dichroic,
giving a dispersion of 0.61\AA~pixel$^{-1}$ and resolution of $\sim$4\AA.

Bias subtraction and flat-fielding
were performed for UH data, where flat
fields were constructed from images of the twilight sky.
Photometry of Scheila's nucleus was performed using circular apertures
(with background statistics measured in
nearby regions of blank sky to avoid coma contamination) and calibrated using \citet{lan92} standard stars.

For our spectroscopic observations, to avoid saturation from the nucleus
while still maximizing the possibility of detecting gas emission in the faint dust cloud,
we used a north-south slit orientation, offset $0\farcs6$
east of the object's photocenter.
North-south dithers (by 30$''$) were employed to sample the sky background.
We obtained two 900s exposures for a total integration time of 1800s.
Bias subtraction, flat-fielding,
and wavelength calibration
were performed using Low-Redux\footnote{Developed by J.\ X.\ Prochaska
and available at http://www.ucolick.org/$\sim$xavier/LowRedux/} software.
Object identification, extraction, and flux calibration were performed using the Image
Reduction and Analysis Facility (IRAF) software package.

\section{RESULTS \& ANALYSIS\label{results}}

\subsection{Photometric Analysis\label{photresults}}

We find a mean $V$-band magnitude for Scheila's nucleus of $m_V=14.22\pm0.01$~mag inside a
photometry aperture $3\farcs0$ in radius (Table~\ref{photom}).
This measurement represents a photometric excess of $\Delta m_V=0.37$~mag from the quiescent magnitude ($m_V=14.59$~mag)
predicted by Scheila's IAU phase law \citep[$H_V=8.84\pm0.04$, $G=0.076\pm0.06$;][]{war10}, which significantly exceeds the
amplitude of its measured rotational lightcurve \citep[0.09~mag;][]{war06},
indicating the presence of unresolved near-nucleus coma equivalent to $\sim40$\% of the
nucleus scattering cross-section inside the photometry aperture.

\begin{figure}
\plotone{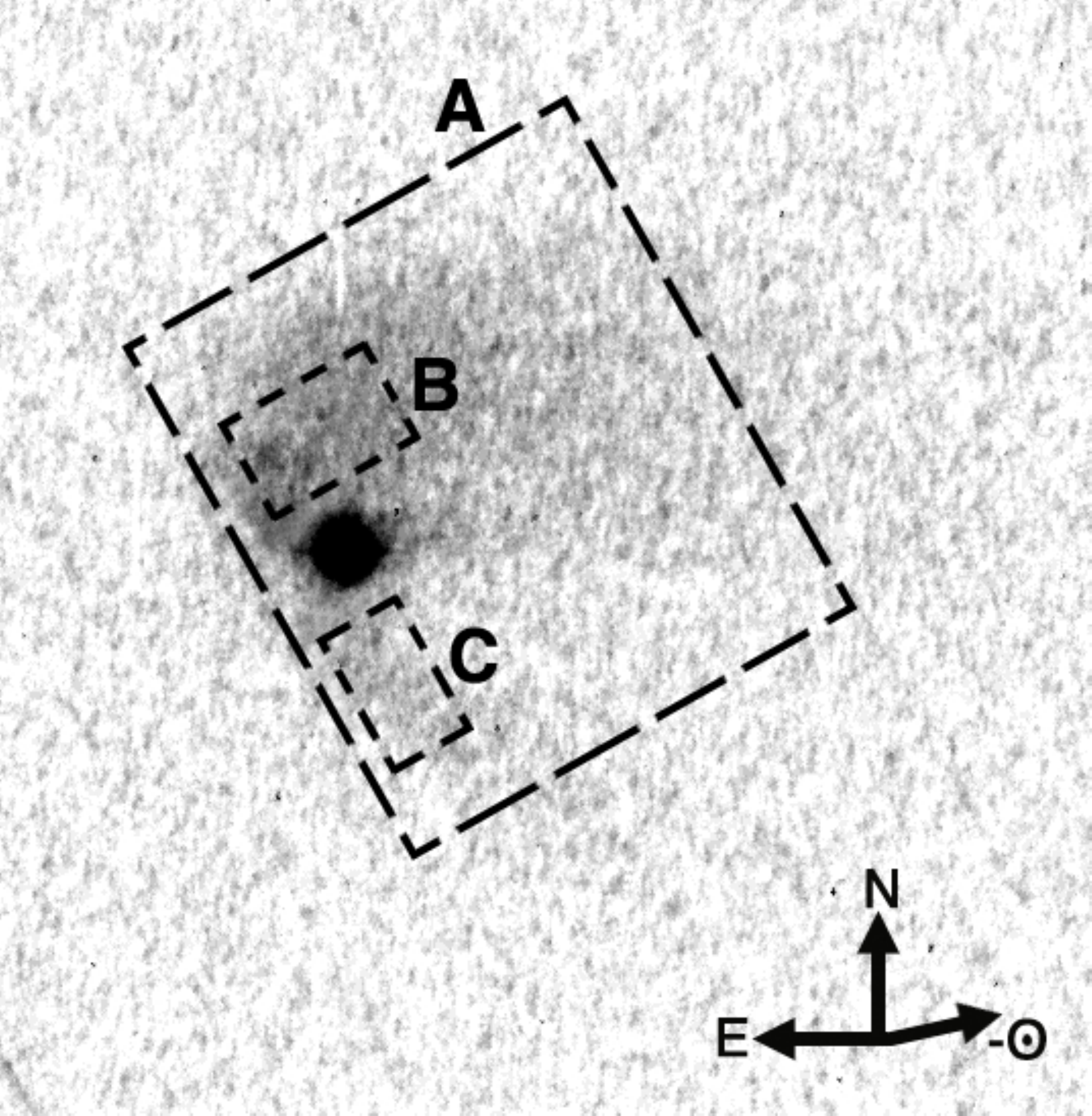}
\caption{\small Composite $R$-band image of Scheila constructed from data obtained using the UH~2.2~m telescope
on 2010 December 12 and comprising data equivalent to 2230~s in total exposure time. 
Rectangular photometry apertures used to measure dust fluxes are marked, where aperture ``A'' is $53''\times62''$ in size,
``B'' is $17\farcs5\times11\farcs0$, and ``C'' is $9\farcs2\times15\farcs9$.
}
\label{scheila_image}
\end{figure}

In a rectangular aperture enclosing the entire dust cloud (``A'' in Figure~\ref{scheila_image}), we find a total
magnitude of $m_V=13.73\pm0.02$~mag, corresponding to a photometric excess of $\Delta m_V=0.86$~mag.
The size of this rectangular aperture is chosen to encompass the entire dust cloud as observed in $V$-band before it
becomes visibly indistinguishable from the background sky.  However, we see in Figure~\ref{scheila_image} that the dust cloud extends
somewhat farther when observed in $R$-band.  Measuring the dust cloud's $R$-band flux using a larger aperture
and comparing to the flux measured using the same aperture used for our $V$-band data, we determine that
the $V$-band flux of the dust cloud may be underestimated by $\sim10$\%, and so for the purposes of estimating the total mass
of Scheila's dust cloud, apply this offset to
obtain aperture-corrected values of $m_V=13.63\pm0.02$~mag and $\Delta m_V=0.96$~mag.
Using the nucleus for calibration, we find a total dust
scattering cross section of $A_d = (1.4\pm0.3)\times10^4$~km$^2$ ($\sim140$\% of the nucleus cross-section).
Assuming a typical bulk density of $\rho=1640$~kg~m$^{-3}$ \citep[cf.\ the Tagish Lake meteorite;][]{hil06}
and an average particle radius of $a_d=1\times10^{-6}$~m, we can estimate the total dust mass, $M_d$, of the cloud using
\begin{equation}
M_d = {4\over3}A_d\rho a_d
\end{equation}
Inserting the scattering cross-section we measure for Scheila's dust cloud, we thus obtain an approximate dust mass
of $M_d\sim3\times10^7$~kg.  This estimate is of course based on numerous assumptions, particularly average
dust grain size and bulk density, as specified above, and is only computed to provide approximate physical context
to the photometric excess that was measured.
For reference, employing the different grain size distribution and density
assumptions made by \citet{jew11} ($\rho=2000$~kg~m$^{-3}$; $a_d=1\times10^{-6}$~m) and
\citet{bod11} ($\rho=2500$~kg~m$^{-3}$; $a_d=1\times10^{-4}$~m), we would instead obtain
$M_d\sim4\times10^7$~kg and $M_d\sim5\times10^9$~kg, respectively.

For comparison, \citet{lar10} measured $\Delta m_V=1.24$~mag on 2010 December 3,
while \citet{bod11} measured $\Delta m_V=0.66$~mag on 2010 December 14-15 and
\citet{jew11} measured $\Delta m_V=1.26$~mag on 2010 December 28 and $\Delta m_V=1.00$~mag
on 2011 January 5.
However, given the different observational and instrumental circumstances involved
(especially given the technical difficulty of observing the extremely bright nucleus and
comparatively extremely faint dust cloud simultaneously),
with the exception of the decline between the two measurements by \citet{jew11},
we do not consider any dust mass fluctuations implied by comparing these
disparate data sets to be reliable.

$V-R$ color measurements (Table~\ref{photom}) of the dust-contaminated nucleus indicate that it is
redder than the Sun, with the entire dust cloud with nucleus flux subtracted
(``Dust$_{\rm A}$'' in Table~\ref{photom}) having an effectively identical red color,
similar to colors measured for other active comets and active Centaurs \citep{bod11,jew09b}.
Assuming that the dust contaminating our nucleus photometry has a scattering cross-section 40\% larger
than that of the nucleus (estimated above) and the same color as the dust cloud as a whole,
we find a dust-subtracted color for the nucleus of $V-R\approx0.44$~mag, consistent with a
D-type spectral classification \citep{for07}.
To test whether the northern and southern plumes of Scheila's dust cloud exhibit any compositional differences, we measure
their colors individually (``Dust$_{\rm B}$'' and ``Dust$_{\rm C}$''), but within estimated
uncertainties, find no significant color differences.

\subsection{Spectroscopic Analysis\label{specresults}}

\subsubsection{Gas Emission Search}

\begin{figure}
\begin{center}
\includegraphics[width=3.6in]{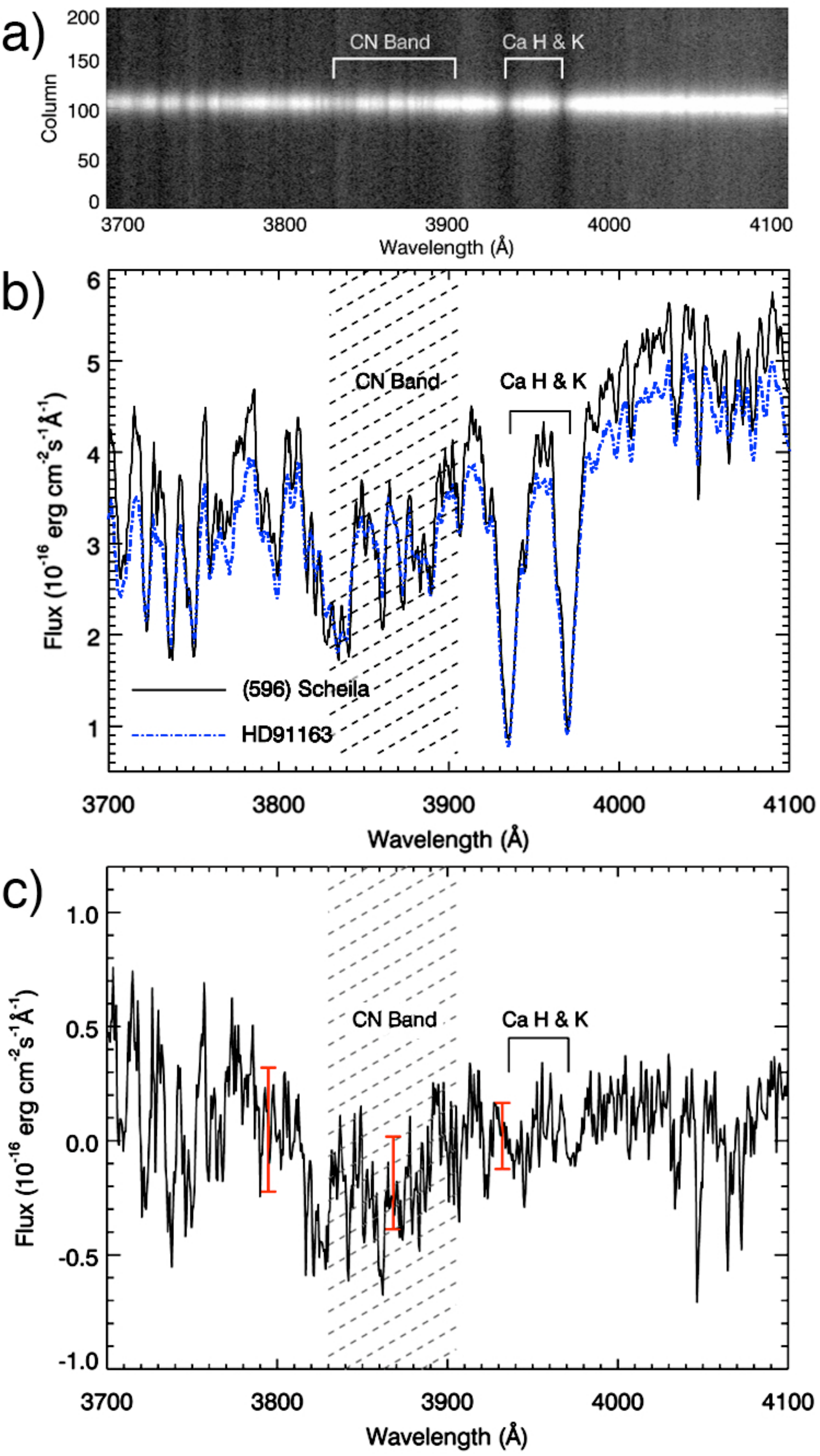}
\end{center}
\caption{(a) Long slit spectral image of Scheila, taken on UT 2010 December 17 using Keck I. 
The wavelength range (3830\AA\ $ < \lambda < $ 3905\AA) where CN emission is expected is marked. Sky emission lines are visible along the spatial direction (the vertical axis) and the solar Ca H and K absorption lines are marked.
(b) Sky background-subtracted and flux-calibrated spectrum of Scheila (solid black line). The scaled spectrum of the solar analog, G2V star HD 91163, is marked as a blue dashed line. The shaded region indicates where CN emission is expected. (c) Spectrum of Scheila with the underlying solar-type continuum subtracted. Red error bars show the 1$\sigma$ uncertainties in the three wavelength regions discussed in the text.}
\label{keck_obs}
\end{figure}

The most sensitive probe of sublimating gas in a comet is CN emission at 3880\AA.
We show the spectral image of Scheila from $3700{\rm \AA}$ to $4100{\rm \AA}$ in
Figure~\ref{keck_obs}a. In this image, the horizontal continuum
corresponds to reflected light from the nucleus.
OH sky emission lines are visible as vertical bands, and
dark Fraunhofer lines in the Solar spectrum and prominent
Ca H (3933\AA) and K (3966\AA) absorption lines are also visible.
In a two-dimensional spectral image, the intensity, $I_e$, of a cometary emission lines should be highest near
the center of the continuum and gradually decrease with increasing distance
from the continuum, moving in the spatial direction. No spectral features near 3880${\rm\AA}$ exhibit
such behavior, and we therefore conclude that no gas is detected. 

We also search for CN emission in a one-dimensional spectrum extracted from the spectral image
using a $1\farcs0\times8\farcs1$ rectangular aperture
centered on the continuum.
Sky background was
measured and subtracted using flanking regions 10$''$ to
16$''$ from the nucleus.
Calibration was performed using a nearby flux standard star and a
solar analog star (Fig.~\ref{keck_obs}b).
Shaded regions in Figures~\ref{keck_obs}b and \ref{keck_obs}c indicate where
CN emission is expected, but we again find no evidence of
emission, consistent with work by
\citet{bod11}, \citet{how11}, and \citet{jeh11}.

\subsubsection{Gas Production Rate Computations}

To estimate Scheila's CN production rate, we remove the continuum using
a scaled solar analog spectrum
(Fig.~\ref{keck_obs}c), which should then leave only gas emission.
Standard errors in three wavelength regions in the residual spectrum
(3760\AA--3830\AA; 3830\AA--3900\AA, where CN emission is expected; and 3980\AA--4050\AA; each 70\AA~in width) are 
$2.7\times10^{-17}$, $2.0\times10^{-17}$, and $1.8\times10^{-17}~{\rm erg~cm}^{-2}$~s$^{-1}$~\AA$^{-1}$, respectively
(Fig.~\ref{keck_obs}c).
We choose
$2.7\times10^{-17}$~erg~cm$^{-2}$~s$^{-1}$~\AA$^{-1}$ as a conservative estimate of
the uncertainty in the CN band.
Since 9P/Tempel~1 observations \citep{mee11}
utilized the same instrumental settings used here, we assume that any CN band in
Scheila's spectrum will have the same profile as in 
9P's spectrum. We therefore find a peak value of any CN band of
$3\times(2.7\times10^{-17}$~erg~cm$^{-2}$~s$^{-1}$~\AA$^{-1}$), or $\sim8\times10^{-17}$~erg~cm$^{-2}$~s$^{-1}$~\AA$^{-1}$. 

We calculate the integrated CN band flux, $f_{\rm CN}$, by summing the emission
flux in the shaded region, obtaining $f_{\rm CN}=8.8\times10^{-16}$~erg~cm$^{-2}$~s$^{-1}$.
We then convert $f_{\rm CN}$ to the total number of CN molecules, $N_{\rm CN}$, using
\begin{equation}
L_{\rm CN} = 4\pi\Delta^2f_{\rm CN}
\end{equation}
\begin{equation}
\log N_{\rm CN}= \log L_{\rm CN} + 2.0\log r_{h} - \log g(l)
\end{equation}
where $\Delta$ and $r_h$ are
in cm and AU, respectively, and $g(l)$ is the resonance fluorescence efficiency,
which describes the number of photons scattered per second per radical, in erg~s$^{-1}$~molecule$^{-1}$.
During our observations, Scheila had a
radial velocity of $\dot{r}_h=2.5$~km~s$^{-1}$, for which $g(l,r_h)=2.7\times10^{-13}$~erg~s$^{-1}$~molecule$^{-1}$
when the Swings effect is taken into account \citep{sch10}.
Substituting $f_{\rm CN}=8.8\times10^{-16}$~erg~cm$^{-2}$~s$^{-1}$, we obtain
$N_{\rm CN}=5.33\times10^{26}$.

A simple \citet{has57} model is used to derive the CN production rate, $Q_{\rm CN}$,
from $N_{\rm CN}$, assuming isotropic outgassing, constant radial expansion of the
gas coma, and a 2-step exponential decay process. We use $l_{p}=1.3\times10^{4}$ and $l_{p}=2.2\times10^{5}$
as the effective Haser scale lengths at $r_{h}=1$~AU
\citep{ahe95}, and adopt a gas velocity of
\begin{equation}
v = 1.112 \cdot r_{h}^{-0.4} ~{\rm km~s}^{-1}
\end{equation}
\citep{biv97}.  Integrating the computed spatial column density model
over a rectangle 1800 km $\times$ 14480 km (the physical size of the
extraction aperture in the sky), we find $Q_{\rm CN} < 9\times10^{23}$~s$^{-1}$. Taking average ratios of species in
previously observed comets
($\log [{Q_{\rm CN}}/{Q_{\rm OH}}]=-2.5$; ${Q_{\rm OH}}/{Q_{\rm H_{2}O}} = 90$\%)
\citep{ahe95},
we estimate a water production rate of $Q_{\rm H_{2}O} < 10^{27}$~s$^{-1}$.
However, given the uncertainties involved
in assuming ratios of comet species measured at much closer heliocentric distances
remain unchanged for an object in the main asteroid belt, we
regard this estimate to be precise, at best, to an order of magnitude.

\subsection{Dynamical Analysis\label{dynresults}}

To gain a more complete understanding of the circumstances surrounding Scheila's unusual outburst,
we also consider various aspects of the object's dynamical nature.  Specifically, we consider its likely origin and
whether it belongs to an asteroid family.

\begin{figure}
\plotone{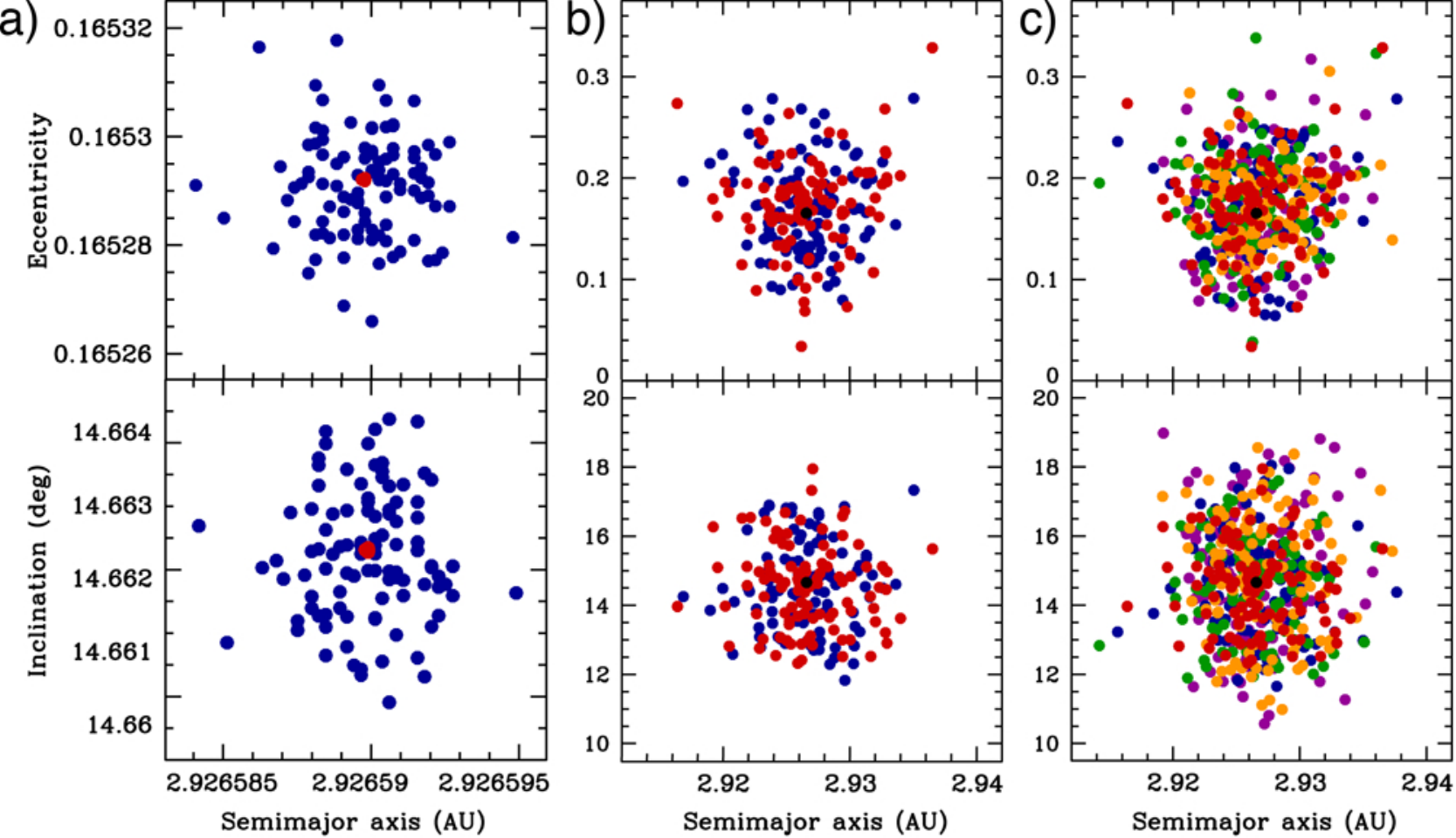}
\caption{Plots of semimajor axis versus eccentricity (top) and inclination (bottom) for (a) the initial orbital elements and (b) final orbital elements
after a 100 Myr dynamical integration
of 100 1-$\sigma$ Scheila test particles (red dots) and 100 100-$\sigma$ test particles (blue dots),
and (c) the orbital elements of the same 100 1$\sigma$ test particles shown at the start of our simulation (overlapping
black dots at the center of each panel) and 100 1-$\sigma$ test particles after 20 Myr (purple), 40 Myr (blue), 60 Myr (green), 80 Myr
(orange), and 100 Myr (red).
}
\label{scheila_sims}
\end{figure}

To address the first issue, an effort motivated by the possibility that cometary objects in the asteroid
belt may not necessarily originate where they are currently seen \citep[cf.\ P/2008 R1 (Garradd);][]{jew09},
we perform numerical simulations to assess Scheila's dynamical stability.
We generate two sets of 100 test particles with Gaussian distributions
in orbital element space, centered on Scheila's JPL-tabulated osculating orbital elements,
where the two sets are characterized by $\sigma$ values equal to 1$\times$ and 100$\times$ the JPL-tabulated uncertainties
for each orbital element (Fig.~\ref{scheila_sims}a).  We then use the N-body integration package, Mercury \citep{cha99}, to integrate
the orbit of each test particle forward in time for 100 Myr.

In three runs
using different randomly generated sets of test particles, no objects escape from the asteroid belt,
indicating that Scheila is dynamically stable over this time period, and is likely not a recent arrival
from elsewhere in the main belt or the outer solar system.
We note that despite the 100-fold difference in their initial dispersions,
objects from both the 1-$\sigma$ and 100-$\sigma$ sets of test particles
diverge to occupy similar regions of orbital element space (Fig.~\ref{scheila_sims}b).
This divergence occurs quickly (within 10$^4$ years)
and then remains approximately constant for the 100 Myr test period (Fig.~\ref{scheila_sims}c),
with all objects remaining roughly confined to 
$2.919<a~{\rm (AU)} <2.936$, $0.06<e<0.28$, and $11.5\degr<i<18.5\degr$.
Objects as large as Scheila in the main asteroid belt are often simply assumed to be
dynamically stable, of course, but the example of the Centaur (2060) Chiron shows that 
it is possible for similarly large objects to occupy dynamically unstable orbits \citep[{\it e.g.},][]{nak93}.
As such, given Scheila's unusual comet-like outburst, it is useful to have
explicit confirmation (provided by our simulations) that the object is in fact dynamically stable,
and is therefore unlikely to be a recently-arrived interloper from elsewhere in the solar system.

Next, given the suggestion that MBCs might be preferentially
found among members of young asteroid families due to
their increased likelihood of having fresh near-surface ice
compared to undisrupted asteroids \citep{hsi09b}, we perform a
hierarchical clustering analysis \citep{zap90,zap94}
on proper elements for $\sim$135,000 asteroids from the AstDys website
\citep[http://hamilton.dm.unipi.it/astdys/;][]{kne03b} to search for asteroids
dynamically related to Scheila.  We find only six asteroids
within a cutoff value of $\delta v'=120$~m~s$^{-1}$, where
commonly recognized families have tens to thousands of dynamically related
members, usually within much smaller cutoff values, as shown by \citet{nes05},
despite the fact that their analysis was performed when far fewer asteroids were
known compared to the present day.
Such a small number of related asteroids indicates that Scheila likely does not
belong to a family.

This conclusion is consistent with Scheila's size:  objects
$\sim100$~km in diameter are predicted to have collisional destruction timescales longer than
the age of the solar system \citep{far98,bot05}.  Thus, any such body observed
today is likely to have a primordial origin.
The Vesta family, however, shows
that even extremely large asteroids can be associated with families, specifically
if those families were produced by large but non-catastrophic cratering impacts
\citep{bin93,tho97}.  While such events will leave parent asteroids mostly intact, they would
still excavate large amounts of surface material.  If a parent asteroid happens to contain ice reservoirs
preserved deep beneath its surface, by removing much (if not all) of the surface material
previously insulating that ice against sublimation by solar heating, large cratering impacts could
render that body far more susceptible to subsequent activating impacts by smaller asteroids,
such as those hypothesized to have triggered activity in other MBCs \citep{hsi06}.
The insignificant number of asteroids found to be dynamically related to Scheila indicates, however,
that not only is Scheila unlikely to be a recently-produced fragment of a larger asteroid, it is also
unlikely to have experienced a large (fragment-producing) but non-catastrophic cratering impact in the
recent past, and so we conclude that its likelihood of having recently exposed preserved icy material from its deep
interior ({\it i.e.}, several km or more below its surface) is low.

\section{DISCUSSION\label{discussion}}

\subsection{Scheila's Physical Nature}

\citet{yan11} found that Scheila's near-infrared spectrum (from 0.8 to 4.0~$\mu$m) exhibits a
consistent red slope, has no apparent absorption features, and generally resembles spectra of D-type
asteroids.  Using an intimate mixing model incorporating water ice, amorphous carbon, and iron-rich
pyroxene, they were able to reproduce Scheila's spectrum except for a clearly visible water absorption feature
that is present in the synthetic spectrum but not the observed spectrum.
Despite the absence of unambiguous evidence of water ice in these observations, given the signal-to-noise
ratio of the data, \citet{yan11} concluded that the presence of water ice on Scheila's surface could not
be excluded to a level of a few percent.
They further noted that the similarity between Scheila and D-type asteroids,
which in turn have been noted for their similarities to classical cometary nuclei from the outer solar system
\citep{fit94}, suggests that Scheila could contain preserved ice deep within its interior \citep[{\it e.g.},][]{jon90}.

For comparison, two other MBCs, 133P/Elst-Pizarro and 176P/LINEAR, have been identified as B-type or
F-type asteroids \citep{bag10,lic11b}, where both types are subgroups of C-type asteroids.
Interestingly, these spectral classifications mean that, unlike Scheila, they are more physically
similar to nearby main-belt asteroids than to classical comets \citep{lic11b}, even though we believe the active episodes for
these objects were actually sublimation-driven in nature, while Scheila's was not.
The nuclei of MBCs 238P/Read, P/Garradd, and P/2010 R2 (La Sagra), as well as that of
P/2010 A2, have yet to be characterized spectroscopically due to the
small sizes of 238P and P/2010 A2 \citep{hsi11b,jew10} and significant cometary
activity at the time of all currently published observations of P/Garradd and P/La Sagra \citep{jew09,mor11a}.

As discussed above (Section \ref{intro}), direct spectroscopic detections of gas emission in MBCs have been elusive
\citep[{\it e.g.},][]{jew09,lic11b}.  Successful unambiguous detections of ice have been likewise difficult to obtain.
No spectroscopic evidence of exposed water ice has been found on any of
the MBCs \citep[{\it e.g.},][]{rou11}.  However, \citet{hsi06} hypothesize
that cometary activity for these objects is being driven by small areas of exposed subsurface ice
\citep[hundreds of square meters on km-sized bodies; cf.][]{hsi04,hsi09b}.  If that approximate ratio of
active to inactive surface material is correct, we would actually expect unresolved disk-integrated reflectance
spectroscopy to be of limited use without sensitivity levels to one part in $\sim10^4$ or better.

While this sensitivity level is beyond the reach of current Earth-bound facilities for
the km-scale (and smaller) MBCs at the distance of the asteroid belt,
\citet{riv10} and \citet{cam10} have reported water ice detections (corresponding to much larger
surface coverage than predicted for the MBCs)
for the 100-km-scale asteroid (24) Themis, which belongs to the same Themis asteroid family 
that also contains 133P and 176P \citep{hsi06}.  A similar absorption feature has also been detected
on outer belt asteroid (65) Cybele \citep{lic11a}.
The attribution of the absorption feature observed
by these groups to water ice has been challenged by \citet{bec11} though, who suggest that the absorption
feature can be equally well explained by the non-volatile mineral goethite.
\citet{riv10} and \citet{cam10} acknowledge the thermal instability of water ice that they claim
to detect and propose various scenarios how such surface ice could be maintained.
However, none of these scenarios has yet been confirmed to actually plausibly account for a
widespread, long-lived surface layer of water ice as implied by their observations.
We further note that no outgassing or dust emission has ever been observed for (24) Themis
\citep[also noted by][]{riv10}, and as such, the connection between ice on main-belt asteroids
and the activity of MBCs remains unsubstantiated.

In summary, while the hypothesis that activity in MBCs is sublimation-driven is supported by indirect
evidence such as numerical modeling results and observations showing recurrent activity
\citep[{\it e.g.},][]{hsi09,hsi10b,hsi11b}, it remains unsubstantiated by the unambiguous detection
of either ice or sublimation products.
As such, if the non-detections of ice or gas emission via spectroscopy currently do not
preclude the designation of an object as an MBC, the same non-detections of ice and gas emission
for Scheila \citep[Section \ref{specresults};][]{how11,jeh11,bod11,yan11}
necessarily must also be considered inadequate criteria for concluding that the object is a disrupted asteroid,
at least on their own.
We must also consider other evidence.

\subsection{Morphology Analysis and Outburst Scenarios\label{morphology}}

The most obvious aspect of Scheila's dust cloud that should bear clues as to its origin is its
unusual morphology.  Unlike many other comets which typically exhibit a single tail, often
pointed in the antisolar direction, and a coma, Scheila exhibits two distinct curved dust plumes extending
to the North and the South (Figure~\ref{scheila_image}).  Deeper and higher resolution imagery also shows
a faint westward-pointing dust ``spike'' \citep{jew11}.  The two main dust plumes
are not easily explained by the standard cometary dust ejection paradigm where grains
are ejected isotropically and subsequently follow well-defined syndyne and synchrone curves \citep[{\it cf}.][]{fin68},
and therefore suggest the action of a more unusual ejection mechanism.

\citet{ish11b} present numerical modeling results that
show that the cloud's morphology can be accounted for by a hollow cone of dust
\citep[as expected from an impact; {\it e.g.},][]{ric07} ejected Sunward and then turned back by
radiation pressure.  The hollowness of the cone is suggested by the dust cloud's apparent limb
brightening ({\it i.e.}, Scheila's northern and southern plumes), consistent with the greater optical depth expected
of such a structure along its edges.  A solid cone of ejected dust (as expected
from a sublimation-driven jet) similarly pushed back would instead exhibit central brightening due to
greater optical depth in the jet's core, and not appear to exhibit multiple plumes.  The disparate strengths
of the northern and southern plumes may indicate an oblique angle of incidence for the impact, with more
dust expected downrange of the inbound impactor \citep[cf.][]{ish11b}.

While the scenario modeled by \citet{ish11b} plausibly accounts for the appearance of multiple dust plumes
with a single impact, if Scheila's dust cloud is produced by a sublimation-driven process, any scenario that similarly accounts for
multiple observed dust plumes likely requires multiple active sites.  If these active sites are collisionally
excavated, multiple impacts would have to have occurred on timescales shorter than their depletion timescales.
This scenario is unlikely, however, given the typically low rate of impacts on any single body in the asteroid
belt \citep[e.g.,][]{far92} and the expected short depletion timescales for surface volatiles on main-belt
asteroids \citep{hsi09b}.  Multiple active sites could be possible if near-surface ice is abundant, triggering
sublimation at multiple points via thermal stresses. This scenario is contradicted though by Scheila's history
of observed inactivity until 2010, and the fact that it first exhibited a comet-like dust cloud at $\nu\sim240\degr$, well before
perihelion, when the surface was receiving only $\sim60$\% of the solar flux that it would at perihelion.
If Scheila were particularly icy, we would expect to have observed past activity and at times of maximum solar
heating.  We therefore find any scenario invoking multiple sublimation sites to be implausible.

One possible scenario which could explain the observed dust morphology via a sublimation-driven
phenomenon would be one in which directed dust emission (i.e., a jet) is being produced from a single active site
near the rotational pole of the object, which in turn is directed approximately Sunward.  Such a situation would
mimic the effect of a hollow cone of material (i.e., the rotating jet) being ejected Sunward, where the cone
would then be expected to be pushed back by radiation pressure as in the impact-driven model described by
\citet{ish11b}.  Asymmetry in the density of material forming the cone could then be caused by diurnal effects
where sublimation, and therefore dust emission, increases in intensity when the active site is at its closest to
the subsolar point and receives maximum Solar heating, and decreases when the active site is farthest from the
subsolar point and it receives somewhat diminished Solar heating, assuming the rotational pole is not directed
exactly at the Sun.

Given Scheila's rotation period of 15.848~hr \citep{war06,ish11a}, this scenario would require Scheila to have thermal
inertia low enough
for the temperature of the active site to vary enough during each
rotation such that the dust production rate of the jet visibly varies as the active site rotates into and out of
direct Solar heating.  This scenario would also require a somewhat fortuitous rotation pole orientation where the
angle between the pole and the Sun-object vector is large enough to create significant diurnal temperature variations
over the course of a single rotation, but small enough that the projected appearance of a Sunward-pointing hollow cone of ejected
material that is then pushed back by radiation pressure is not lost.  The thermal and numerical dust modeling required
to test this scenario is beyond the scope of this paper.  We note, however, that the oblique impact
scenario described by \citet{ish11b} is far simpler than one involving a rotating, diurnally-varying cometary jet oriented
exactly in such a way that it mimics the asymmetric Sunward-directed hollow cone of ejected material that otherwise arises
naturally in the impact scenario.
As such, in the absence of compelling evidence that sublimation must be taking place,
we favor that impact scenario as the most plausible explanation for Scheila's unusual dust emission,
and therefore consider it to be the second disrupted asteroid to be discovered, after P/2010 A2.

\subsection{MBCs vs.\ Disrupted Asteroids}

New and upcoming all-sky survey facilities like the Panoramic Survey Telescope And Rapid Response System \citep[Pan-STARRS;][]{kai10}
and the Large Synoptic Survey Telescope \citep[LSST;][]{jon09}, as well as long-running existing surveys like LINEAR, Spacewatch,
and the Catalina Sky Survey, are expected to discover many more comet-like objects in the future, some of which
may be found to orbit in the main asteroid belt.  A natural question to pose, then, is how can we distinguish with any degree of
certainty which of these objects are MBCs, where comet-like activity is due to sublimation, and which are disrupted asteroids,
where comet-like activity is due to a recent impact?

Unfortunately, a definitive answer to this question does not yet exist.  For this reason, \citet{jew11} have taken the approach of
designating all comet-like objects with main-belt orbits as MBCs, based on their appearance alone ({\it i.e.}, using
the observational definition of a comet as any solar system object exhibiting mass loss in the form of extended surface brightness features
such as a coma or tail).  Using this definition,
they classify Scheila as an ``impact-activated MBC'', agreeing with this work in terms of physical conclusions, if not in
terminology.  We prefer, however, to retain the physical meaning of a ``comet'' as an active body whose
gas or dust emission is due to the sublimation of volatile ices.  As such, we choose to define MBCs as only those
comet-like objects with main-belt orbits whose activity is sublimation-driven (according to our best judgement), as originally specified in \citet{hsi06},
and objects like P/2010 A2 and Scheila, where dust emission is the direct consequence of an impact, as disrupted asteroids.

Of course, imposing the requirement that activity
must be sublimation-driven for an object to be considered an MBC
implies that we know the underlying cause of apparent activity for every comet-like main-belt
object, which is not necessarily always the case (or indeed ever the case, at least not until the first successful gas detection is made).

From a practical standpoint, the development of an unambiguous method of
distinguishing sublimation-driven dust emission and impact-driven dust emission will
require the discovery and thorough individual investigations of many more examples of both types of objects before
a set of reliable criteria can be developed for each,
meaning that significant additional work is required before such a classification system can be achieved.

In the meantime, we can draw the following preliminary conclusions and observations regarding the differences
between MBCs and disrupted asteroids, based on the currently extremely limited populations of each:
\begin{enumerate}

\item{{\it Long-lived activity is not a reliable indicator of a dust emission event's source.}  All five currently
  recognized MBCs exhibited activity persisting over weeks or months \citep{hsi04,hsi09,hsi11a,jew09,mor11a}, but so did
  the two currently recognized disrupted asteroids \citep{jew10,jew11,sno10}.  Long-lived activity in the five recognized MBCs
  is regarded as an indication of continuous emission over an extended period of time, a characteristic of cometary emission.
  However, similar long-lived ``activity'' ($>$1 month), or rather the persistent presence of a dust cloud,
  in P/2010 A2 and Scheila are explained as a consequence of the impact events on each
  object ejecting large ($\gg1$~$\mu$m) particles, which simply take longer to dissipate than smaller particles \citep{jew10,bod11}.

  Observations showing steadily increasing quantities of dust during long-lived active episodes \citep[{\it e.g.}, for 238P;][]{hsi11b}
  could be better indications of sublimation-driven activity since they cannot simply be explained by persistent
  large dust grains.  Ongoing dust particle production is required.  However, while this can result from the
  actual release of new dust from the nucleus, an increase in visible scattering surface area could also arise
  via fragmentation of existing dust particles, as likely occurred during Comet 17P/Holmes's 2007 outburst \citep{hsi10a}.

  In the case of Scheila, \citet{jew11} argued that observations of rapid fading of the dust cloud (approximately one month after
  its discovery) was an indication of impact-driven emission.
  While the short time between the appearance of Scheila's dust cloud (in December 2010) and this rapid fading (observed in January 2011)
  means that this interpretation is likely correct, we note that decreasing activity (with no earlier observations of increasing activity)
  was also observed for 133P in 2002 \citep{hsi04} and P/Garradd in 2008 \citep{jew09}.  Thus, while increasing activity may suggest cometary emission,
  and decreasing activity may suggest impact-driven emission, neither should be considered conclusive proof of either scenario without
  other supporting evidence.
  }
\item{{\it Dust clouds with unusual morphologies may indicate unusual formation circumstances.}
  As discussed in Section~\ref{morphology} (and references within) for Scheila and by \citet{jew10} and \citet{sno10}
  for P/2010 A2, the unusual structures of their respective dust clouds appear to be indications of non-sublimation-driven
  origins.  In the case of P/2010 A2, an observed gap between the nucleus and the dust tail strongly suggested that a
  dust cloud had been ejected in a single impulsive event and was slowly drifting away from the nucleus, consistent with being
  an ejecta cloud produced by an impact, and inconsistent with cometary dust emission.  
  The implications of Scheila's pair of dust plumes are considered at length in Section~\ref{morphology}.
  While such interpretations can be used to draw preliminary conclusions,
  we caution against relying solely on morphology-based
  determinations of a dust emission event's source.  Multi-tail structures have been observed for comets believed to exhibit
  sublimation-driven activity \citep[{\it e.g.}, 238P and P/La Sagra;][]{hsi09,mor11a} and the abrupt termination of sublimation-driven
  dust emission that could cause an observable gap between a nucleus and a dust tail, while probably uncommon, could be possible given certain
  topographical circumstances ({\it e.g.}, shadowing of isolated active sites).  Detailed numerical dust modeling must be
  employed to ensure that any unusual morphological structures cannot be plausibly explained by both sublimation-driven
  and impact-driven dust emission.
}
\item{{\it Recurrent activity, separated by intervening periods of inactivity, is extremely difficult to explain as the
  consequence of impacts.}  Upon the initial discovery of a comet-like main-belt asteroid, conclusions
  about the nature of its activity have generally been the result of numerical dust modeling \citep{boe96,hsi04,hsi09,hsi11a,mor11a}.
  The underconstrained nature of these modeling efforts, however, means their solutions are usually never unique,
  and simply reflect the most plausible ejection scenarios as judged by the authors of those studies.
  
  In contrast, observations of recurrent activity for a single object leave far less room for ambiguity
  \citep[{\it cf}.\ 133P and 238P;][]{hsi04,hsi10b,hsi11b}.  In those cases, repeated active episodes occurred on timescales of $\sim$5-6~years,
  which correspond closely to the orbital period of each object, and are far shorter than the expected
  timescales of repeated random
  collisions on individual main-belt objects by impactors of appreciable size \citep[{\it e.g.},][]{far92}.
  One might envision a scenario in which a main-belt object routinely encounters
  an overdensity of impactors at certain parts of its orbit, perhaps due to an intersecting cometary debris stream, for example.
  The contrived nature of such a scenario (in which debris streams dense enough to reliably impact 133P and 238P on each orbital passage
  only intersect these specific objects and do not cause comparable events on other asteroids with similar orbits)
  argues strongly against its plausibility though. 
  On the other hand, repeated active episodes are routinely observed for comets
  for which activity is driven by sublimation, and can be naturally explained for the MBCs as the consequence of seasonal
  variations in solar heating of isolated active sites, or perhaps simply from the increase in solar heating experienced by
  each object near perihelion \citep{hsi04,hsi11b}.
}
\end{enumerate}

As discussed above, the sizes of the known populations of both MBCs and disrupted asteroids, from which these generalizations
are drawn, are extremely small, and so the typical caveats associated with small-number statistics certainly apply.
However, for the above reasons, we suggest that repeated activity is the least ambiguous and most reliably obtainable
indicator available at the current time that comet-like activity is
sublimation-driven for a particular object.  While only two of the seven currently known comet-like main-belt objects ({\it i.e.},
both MBCs and disrupted asteroids) have been observed to exhibit recurrent activity to date, we note that the remaining five objects were discovered
as comet-like bodies recently enough that they have not actually yet completed full orbits since their respective discoveries.
As such, continued monitoring of all of these objects to search for recurrent activity will
be important for validating their identification as MBCs or disrupted asteroids.


\section{Summary}

We have conducted a photometric, spectroscopic, and dynamical study of comet-like main-belt asteroid
(596) Scheila, and report the following findings:
\begin{itemize}
\item{Scheila's dust cloud comprises a $V$-band photometric excess of
$\Delta=0.96$~mag over the expected quiescent brightness of the nucleus, giving it a scattering cross-section $\sim$1.4 times
larger than that of the nucleus, and an estimated total mass of $M_D\sim3\times10^7$~kg.
Separate $V-R$ color measurements of each of the two curved plumes of the dust cloud show that they are
both redder than the Sun, with no signficant color differences between the plumes themselves.
}
\item{We find no evidence of CN emission in Scheila's dust cloud
down to the sensitivity level of our spectroscopic observations,
and place upper limits to CN and H$_2$O production rates of
$Q_{\rm CN}=8.91\times10^{23}$~s$^{-1}$ and $Q_{\rm H_2O}\sim10^{27}$~s$^{-1}$.
}
\item{An analysis of Scheila and its dynamical vicinity using numerical simulations
indicate that it is dynamically stable for $>100$~Myr, suggesting that it is
likely native to its current location.  We also find that it does not belong to a dynamical asteroid family of any significance,
meaning that it is unlikely to be the product of a recent fragmentation or cratering event
that could have exposed deeply buried ice.
}
\end{itemize}

Considering the available evidence obtained by us and others, we conclude that Scheila is most likely a disrupted asteroid,
and not an MBC (where dust emission is the result of the
sublimation of ice, as in other comets), in agreement with previous work.  This conclusion is largely founded on consideration of the unusual
morphology of Scheila's dust cloud, coupled with the simplicity of the impact scenario required to reproduce it and the
inability of any physically plausible sublimation-driven scenarios to do likewise.
While the currently known examples
of MBCs and disrupted asteroids suggest that morphological analyses can provide strong indications as to the nature of comet-like
activity in main-belt objects discovered in the future, we caution against relying
on such evidence on its own.  We suggest instead that recurrence (or absence of recurrence) of comet-like
activity is the best indicator available to date of whether an object is likely to be a MBC or a disrupted asteroid,
underscoring the importance of long-term monitoring observations of these objects.

\begin{acknowledgements}
H.H.H. is supported by NASA
through Hubble Fellowship grant HF-51274.01, awarded by the Space Telescope Science Institute,
operated by the Association of Universities for Research in Astronomy, Inc., for NASA, under contract NAS 5-26555.
B.Y. and N.H. acknowledge support through the NASA Astrobiology Institute
under Cooperative Agreement No.\ NNA08DA77A issued through the Office of Space Science.
We thank Carl Hergenrother for first alerting us to Scheila's unusual activity,
Anthony Readhead, Roger W.\ Romani, and Joseph L.\ Richards for obtaining Keck observations for us,
Marc Kassis and Richard Morriarty for technical assistance, David
Nesvorn\'y for providing hierarchical clustering analysis software, David Jewitt,
Masateru Ishiguro, and Iwan Williams for valuable discussions, and an anonymous referee for helpful
comments on this manuscript.
\end{acknowledgements}

\begin{deluxetable}{llcrrcrrrrrr}
\tablewidth{0pt}
\tablecaption{Observation Log\label{obslog}}
\tablecolumns{12}
\tablehead{
\colhead{UT Date} &
 \colhead{Telescope} &
 \colhead{Obs.\tablenotemark{a}} &
 \colhead{$N$\tablenotemark{b}} &
 \colhead{$t$\tablenotemark{c}} &
 \colhead{Filters} &
 \colhead{$\theta_s$\tablenotemark{d}} &
 \colhead{$\nu$\tablenotemark{e}} &
 \colhead{$r_h$\tablenotemark{f}} &
 \colhead{$\Delta$\tablenotemark{g}} &
 \colhead{$\alpha$\tablenotemark{h}} &
 \colhead{$\alpha_{pl}$\tablenotemark{i}}
}
\startdata
2010 Dec 12 & UH~2.2m & Im & 81 & 1770 & $V$ & $0\farcs8$ & 239.6 & 3.107 & 2.529 & 16.4 & -1.0 \\
2010 Dec 12 & UH~2.2m & Im & 97 & 2230 & $R$ & $0\farcs8$ & 239.6 & 3.107 & 2.529 & 16.4 & -1.0 \\
2010 Dec 17 & Keck I  & Sp &  2 & 1800 & --- & $1\farcs0$ & 240.5 & 3.100 & 2.462 & 15.6 & -1.5
\enddata
\tablenotetext{a}{Type of observation (Im: imaging; Sp: spectroscopy)}
\tablenotetext{b}{Number of images}
\tablenotetext{c}{Total effective exposure time in seconds}
\tablenotetext{d}{FWHM seeing in arcsec}
\tablenotetext{e}{True anomaly in arcsec}
\tablenotetext{f}{Heliocentric distance in AU}
\tablenotetext{g}{Geocentric distance in AU}
\tablenotetext{h}{Solar phase angle in degrees}
\tablenotetext{i}{Orbit plane angle (between the observer and object orbit plane as seen from the object) in degrees}
\end{deluxetable}

\begin{deluxetable}{crrr}
\tablewidth{0pt}
\tablecaption{Nucleus and Dust Photometry\label{photom}}
\tablecolumns{4}
\tablehead{
\colhead{} &
 \colhead{$m_V$} &
 \colhead{$m_R$} &
 \colhead{$V-R$}
}
\startdata
Nucleus\tablenotemark{a}     & 14.22$\pm$0.01 & 13.77$\pm$0.01 & 0.45$\pm$0.02 \\
Total$_{\rm A}$\tablenotemark{b} & 13.73$\pm$0.02 & 13.27$\pm$0.02 & 0.45$\pm$0.03 \\
Dust$_{\rm A}$\tablenotemark{c}  & 14.84$\pm$0.02 & 14.36$\pm$0.02 & 0.48$\pm$0.03 \\
Dust$_{\rm B}$\tablenotemark{d}  & 22.79$\pm$0.03 & 22.78$\pm$0.03 & 0.51$\pm$0.04 \\
Dust$_{\rm C}$\tablenotemark{e}  & 23.43$\pm$0.04 & 23.00$\pm$0.04 & 0.43$\pm$0.06
\enddata
\tablenotetext{a}{Photometry of nucleus (including unresolved near-nucleus coma) inside a $3\farcs0$ (radius) circular photometry aperture }
\tablenotetext{b}{Photometry of nucleus and dust inside aperture ``A'' as marked in Figure~\ref{scheila_image}}
\tablenotetext{c}{Nucleus-subtracted photometry of dust inside aperture ``A'' (Fig.~\ref{scheila_image})}
\tablenotetext{d}{Photometry of dust inside aperture ``B'' (Fig.~\ref{scheila_image})}
\tablenotetext{e}{Photometry of dust inside aperture ``C'' (Fig.~\ref{scheila_image})}
\end{deluxetable}

\end{document}